\documentclass[aps,prb,superscriptaddress,showpacs]{revtex4}
\usepackage{graphicx}

\begin{document}
\title{The unfair consequences of equal opportunities: comparing exchange models of wealth distribution}
\author{G. M. Caon}
\author{S. Gon\c{c}alves}
\author{J. R. Iglesias}
\affiliation{Instituto de F\'isica, Universidade Federal do Rio
Grande do Sul, Caixa Postal 15051, 91501-970  Porto Alegre  RS,
Brazil}
\begin{abstract}
Simple agent based exchange models are a commonplace in the study of
wealth distribution of artificial societies. Generally, each agent
is characterized by its wealth and by a risk-aversion factor, and
random exchanges between agents allow for a redistribution of the
wealth. However, the detailed influence of the amount of capital
exchanged has not been fully analyzed yet. Here we present a
comparison of two exchange rules and also a systematic study of the
time evolution of the wealth distribution, its functional
dependence, the Gini coefficient and time correlation functions. In
many cases a stable state is attained, but, interesting, some
particular cases are found in which a very slow dynamics develops.
Finally, we observe that the time evolution and the final wealth
distribution are strongly dependent on the exchange rules in a
nontrivial way.
\end{abstract}
\maketitle
%
\section{Introduction}
Empirical studies of the distribution of income of workers,
companies and countries were first presented, a little more than a
century ago, by Italian economist Vilfredo Pareto. He asserted that
in different countries and times the distribution of income and
wealth follows a power law behaviour, i.e.  the cumulative
probability $P(w)$ of agents whose income is at least $w$ is given
by $P(w) \propto w^{-\alpha}$~\cite{Pareto}. Today, this power law
distribution is known as Pareto distribution, and the exponent
$\alpha$ is named Pareto index. However, recent data indicates that,
even though Pareto's distribution provides a good fit to the
distribution of high range of income, it does not agree with
observed data over the middle and low range of income. For instance,
data from Japan~\cite{souma,nirei}, Italy~\cite{clementi},
India~\cite{sinha1}, the United States of America and the United
Kingdom~\cite{dragu2000,dragu2001a,dragu2001b} are fitted by a
log-normal or Gibbs distribution with a maximum in middle range plus
a power law for the high income strata. The existence of these two
regimes may be justified in a qualitative way by stating that in the
low and middle income class the process of accumulation of wealth is
additive, causing a Gaussian-like distribution, while in the high
income class the wealth grows in a multiplicative way, generating
the power law tail~\cite{nirei}.

Different models of capital exchange among economic agents have been
proposed trying to explain these empirical data. Most of these
models consider an ensemble of interacting economic agents that
exchange a fixed or random amount of a quantity called ``wealth''.
The wealth represents the welfare of the agents. The exact choice of
this quantity is not straightforward. For instance, in the model of
Dragulescu and Yakovenko~\cite{dragu2000} the wealth is associated
with the amount of money a person has available to exchange. Within
this model the amount of money corresponds to a kind of economic
``energy'' that may be exchanged by the agents in a random way and
the resulting wealth distribution is -- unsurprisingly -- a Gibbs
exponential distribution. An exponential distribution but as a
function of the square of the wealth is obtained in a model with
extremal dynamics where some action is taken, at each time step, to
change the wealth of the poorest agent, trying to improve its
economic state~\cite{PIAV2003}. In the case of this last model a
poverty line with finite wealth is also obtained, describing a way
to diminish inequality in real societies~\cite{SI2004}.

Aiming to obtain distributions with power law tails, several methods
have been proposed. Keeping the constraint of wealth conservation a
detailed studied proposition is that each agent saves a fraction
--constant or random-- of their
resources~\cite{chatter1,chatter2,chakra,west,sinha2,IGPVA2003,IGVA2004},
Numerical results, as well as recent analytical
calculations~\cite{cristian06}, indicate that one possible result of
that model is condensation, i.e. concentration of all available
wealth in just one or a few agents. To overcome this situation,
different rules of interaction have been applied, for example
increasing the probability of favoring the poorer agent in a
transaction ~\cite{west,IGVA2004}. However, to our knowledge, there
are few detailed studies comparing the effect of these two
parameters, the risk-aversion parameter and the probability of
favoring the poorer agent~\cite{cristian06}. Besides, most of the
previous works do not consider the time evolution of the system,
moreover many of them do not guarantee that a steady state was
attained indeed.

We present here a systematic study of an agent based model where
exchanges are made by pairs of agents chosen at random, so it is
model with no underlying lattice. Each agent, $i$, is characterized
by a wealth, $w_i$ and a risk-aversion factor $\beta_i$, while in
the exchange there is a probability of favoring the poorer partner
given by~\cite{west,IGVA2004}:
\begin{equation}
\label{eq:sca}
p=\frac{1}{2}+f\times\frac{|w_{i}(t)-w_{j}(t)|}{w_{i}(t)+w_{j}(t)},
\end{equation}
where $f$ is a factor going from $0$ (equal probability for both
agents) to $1/2$ (highest probability of favoring the poorer agent).
Thus, in each interaction the poorer agent has probability $p$ of
earn a quantity $dw$, whereas the richer one has probability $1-p$.
We focus on the choice of the quantity $dw$ transferred from the
loser to the winner. In most of the previous work\cite{IGVA2004} a
kind of {\it fair} lottery is used: the amount of money exchanged
correspond to the minimum stake among the partners: $dw=\min
[(1-\beta_{i})w_{i}(t);(1-\beta_{j})w_{j}(t)]$, so it is the same
amount for both agents. However this equal opportunity rule produces
an evil after-effect: in the $f=0$ case, ``condensation'' occurs:
all available wealth goes to one (or very few) agent, i.e. the Gini
coefficient converges to 1. Also, for $f \neq 0$, even if the poorer
partner has bigger chances of winning, its gains are as negligible
as its own capital, so chances of improving are very
low~\cite{hayes}. For this reason we also investigate an alternative
rule, where $dw$ is just the amount risked by the loser --
$(1-\beta_{j})w_{j}(t)$ -- being $j$ the loser agent. We call this
rule {\it loser} rule. Actually, variations of this {\it loser} rule
have been used in some of the papers quoted
above~\cite{dragu2000,chatter1,chatter2,chakra,sinha2}, but there is
no a good reason why a rich agent will risk more than its poorer
partner. Possible examples are marriage followed by divorce, the
parties do combine their holdings and later divide them, or, in the
corporate world, mergers followed by spin-offs~\cite{hayes}.

In what follows we compare the results between the two rules in
terms of the following quantities that we define thereafter: wealth
distribution $H(w)$, Gini index vs. time, and wealth temporal
correlation function. The wealth distribution is probably the most
important quantity for the global description of a economic system.
$H(w)$ vs. $dw$ gives the fraction of the population that have
wealth between $w$ and $w + dw$.  However, this distribution is
obtained at a given time, both in real situations as well as in
simulations, and in the case of simulations it is important to know
weather the results are stable or not. With this purpose we measure
the Gini coefficient as a function of time. It represents a
practical way to verify the time dependence of the economic
parameters. Finally, we also present the wealth temporal
autocorrelation function. For one side this is another possible
measure of time dependence in a system. Besides, it is more
sensitive because it depends on two times, so aging properties, if
there is any, could be grasped.

\section{Wealth distributions}
All the simulations have been performed for a system of 1000 agents
and the results have been averaged over 1000 samples. Wealth
distributions are evaluated at the {\it final stage}, while at the
initial condition, both wealth ($w_i$) and risk-aversion factor
($\beta_i$) are uniformly distributed in the $[0,1]$ interval.  {\it
Final stage} means not further ---or very small--- changes are
observed as time goes by. Support for that assumption will be
presented in the next section.
\begin{figure}[htb]
\centering{\includegraphics[width=5cm,angle=-90,clip=true]{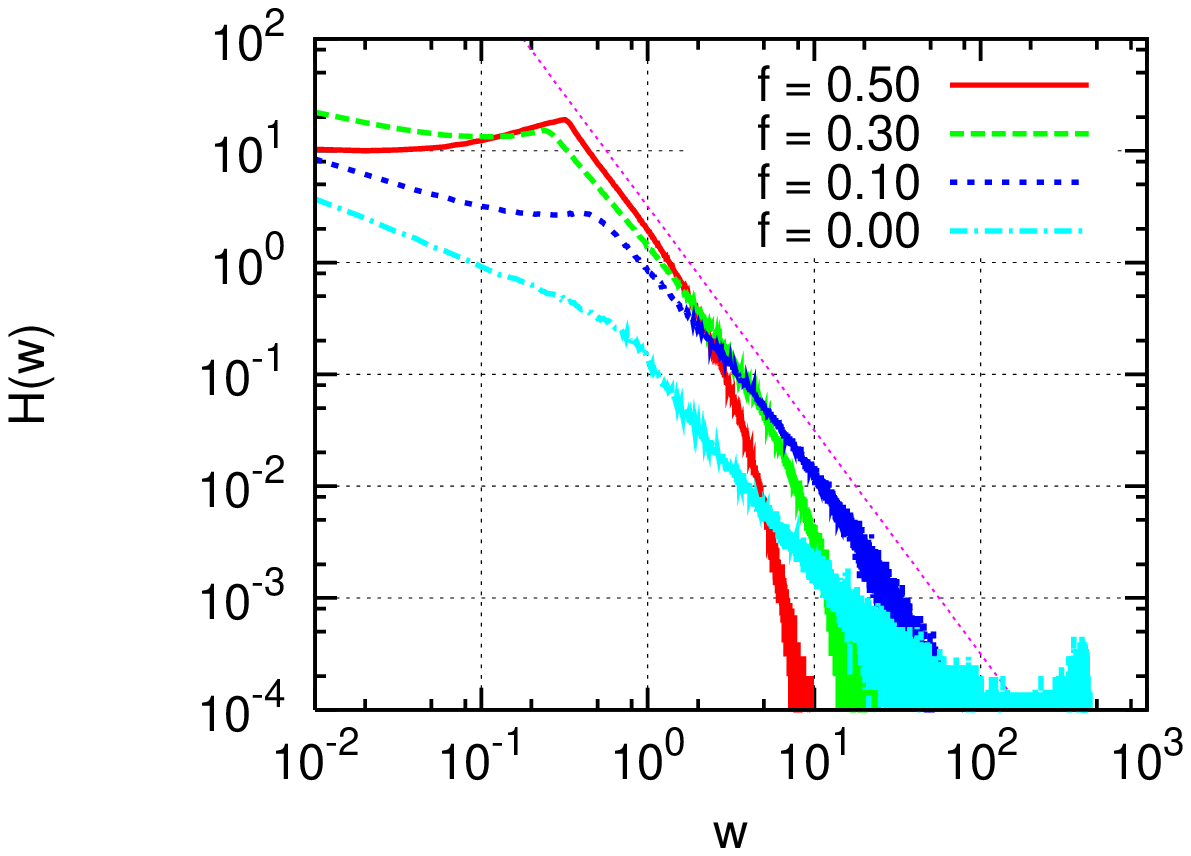}
           \includegraphics[width=5cm,angle=-90,clip=true]{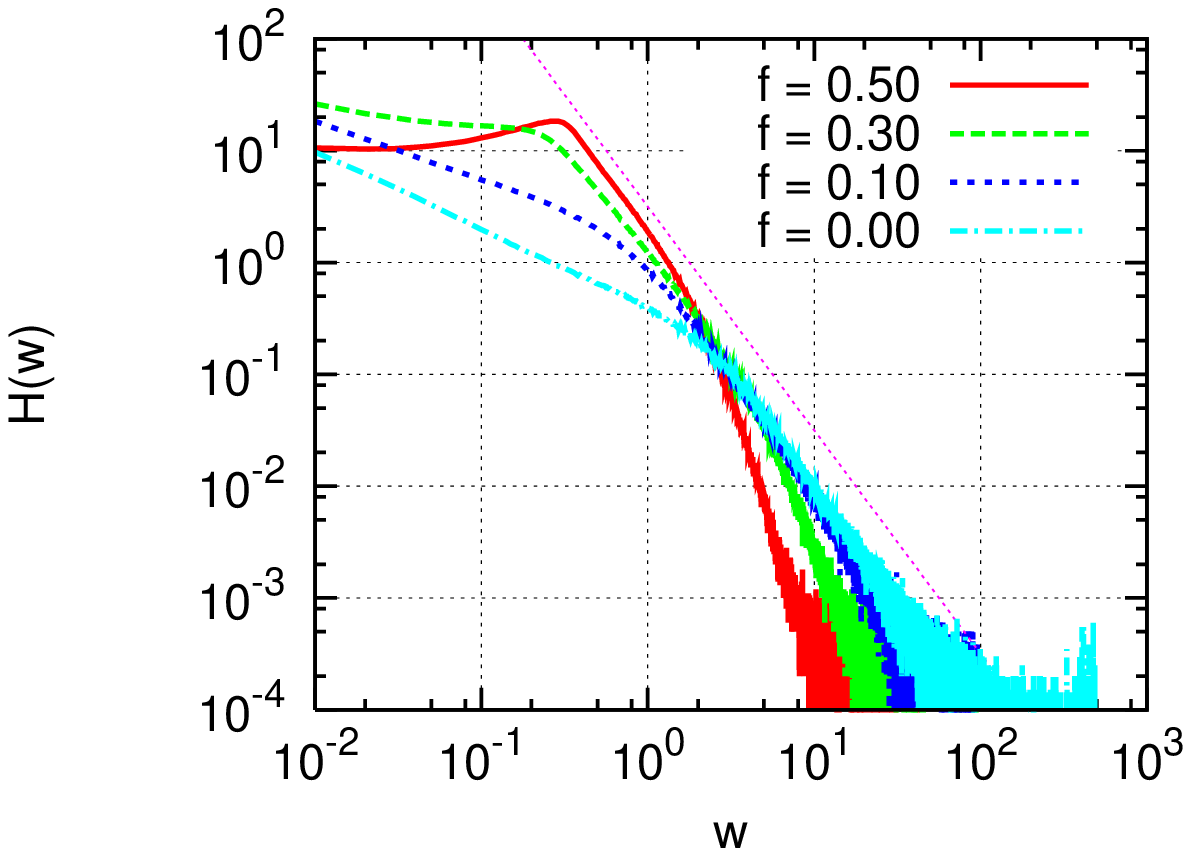}}
\caption{Wealth distribution for several values of the parameter
$f$, with {\it fair} (left) and {\it loser} (right) rule. The
straight line with slope $2$ corresponds to the slope of the $f=0.5$
case, in the intermediate income region, for both rules.}
\label{hist}
\end{figure}
At first both rules seem to be qualitatively similar (one of them, the
{\it fair} rule, was already discussed in ref.~\cite{IGVA2004}). The
main features of the wealth distributions emerging of the present
model are: a) an almost uniform region for very low incomes ($w < 3
\times 10^{-1}$), being more uniform as $f$ is bigger, b) a power law
region extended over less than three decades with an $f$-dependent
exponent. The distinctive feature is in the intermediate income region
which appears less populated, at low $f$ values, in the {\it fair}
rule, when compared to the {\it loser} rule. The straight line
depicted in both graphs, with slope equal to two, is a guide to the
eye, and it follows approximately the slope of the distributions in
the intermediate income region, with $f=0.5$, for both rules. We would
like to emphasize the $f=0$ case, where both distributions exhibit a
power law, but the loser case has a bigger exponent ($-2.17$ compared
to $-1.93$ for the fair case) so indicating a less unequal
distribution.

\section{Gini coefficient} 
\begin{figure}[htb]
\centering{\includegraphics[width=5cm,angle=-90,clip=true]{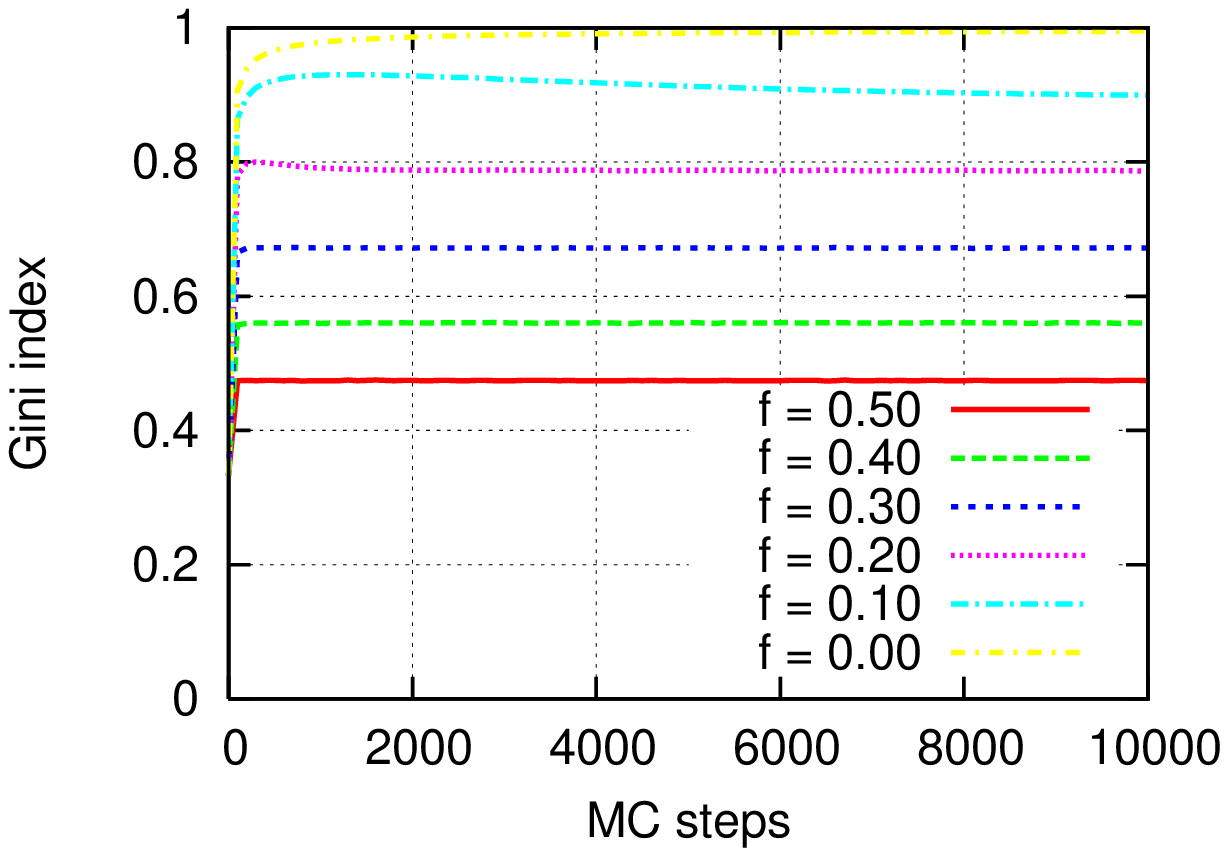}
           \includegraphics[width=5cm,angle=-90,clip=true]{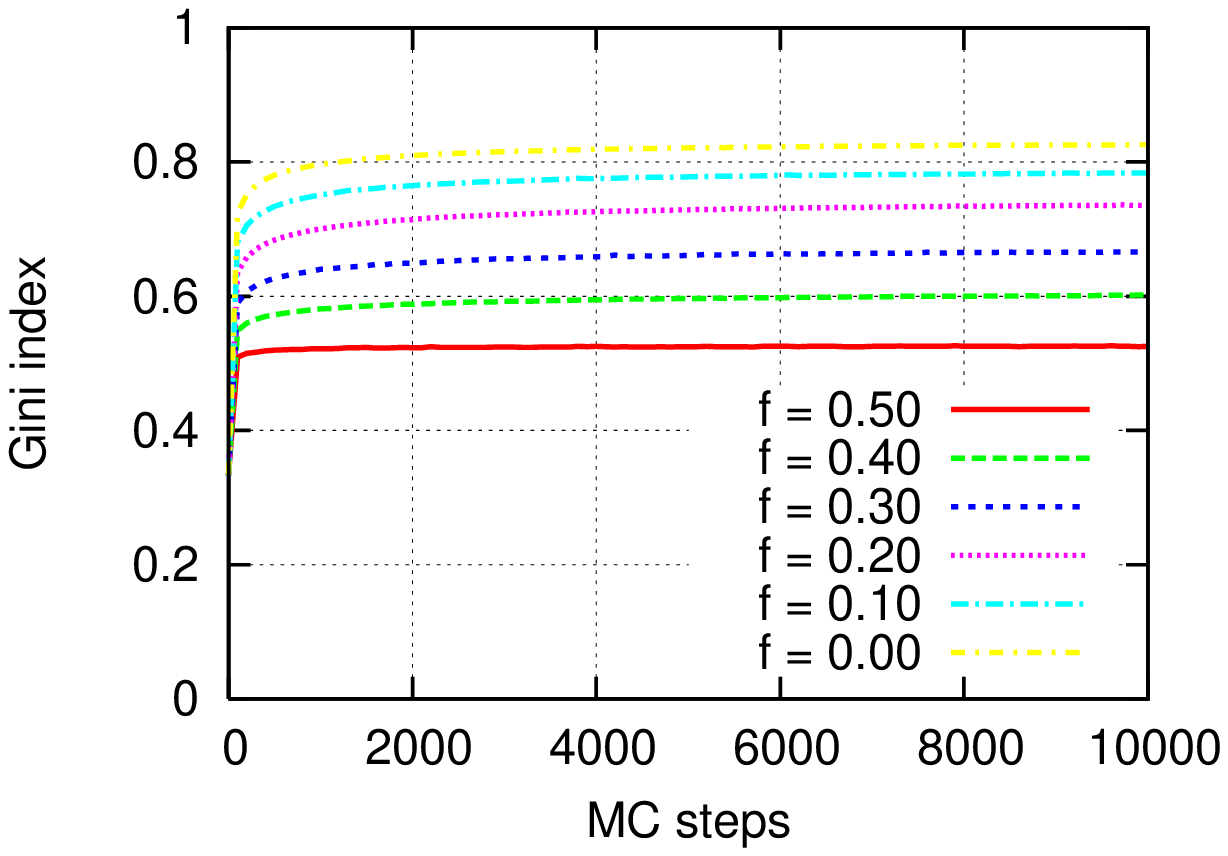}}
\caption{Gini index as a function of time for different values of
$f$, with {\it fair} (left) and {\it loser} (right) rule.}
\label{gini1}
\end{figure}

The Gini coefficient is a measure
of the inequality of a distribution and is often used to measure
income inequality by economists and statistical organizations
because it gives a raw picture of the inequalities with a single
number between 0 and 1. It is defined as the ratio of the area
enclosed by the Lorenz curve of the distribution (or cumulative
distribution function) and the curve of the uniform distribution, to
the area under the uniform distribution\cite{Gini}. In a operational
way we define the Gini coefficient as:
\begin{equation}
G=\frac{1}{2}\frac{\sum_{i,j}|w_i-w_j|}{N \sum_i w_i}
\end{equation}
and it is evident that it varies between 0, which corresponds to
perfect equality (i.e. everyone has the same income), and 1, that
corresponds to perfect inequality (i.e. one person has all the
income, while everyone else has zero income). Here we will use the
Gini coefficient to measure the degree of inequality, but also to
determine the stability of the wealth distribution. With this
objective in mind, we show in Fig.~\ref{gini1} the time evolution of
the Gini index,
where we can observe that the systems converge rather fast to a
fixed value, indicating that, after a transient, the systems arrive
to an almost stable wealth distribution. A further analysis
(measuring the slope) reveals that in some cases stabilization is
not complete, and that a slow dynamics is still present. This point
is confirmed by the analysis of the time correlation function
presented in the next section. The effect of $f$ on the Gini
coefficient is the expected one, the bigger the value of $f$, the
bigger the probability of favoring the poorer agent in each
transaction, and the lower the Gini index. However, the {\it loser}
rule produces, for low values of $f$, a considerable lower Gini
index than the {\it fair} rule, suggesting that an unfair lottery
produces a more equal society than a fair one. Particularly, for
$f=0$ the {\it fair} rule delivers a Gini coefficient equal to $1$
(full condensation of the economy), while the {\it loser} rule
produces a finite Gini coefficient near $0.8$. This is a result that
can induce some second thoughts about the concept of equal
opportunity~\cite{hayes}.

\begin{figure}[htb]
\centering{\includegraphics[width=5cm,angle=-90,clip=true]{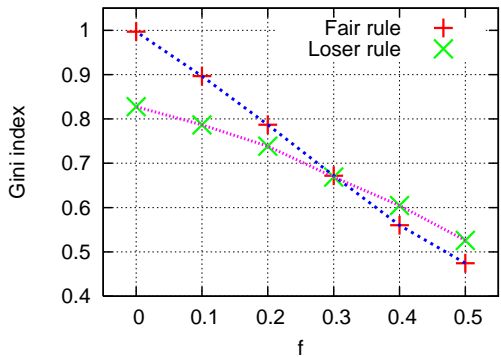}}
\caption{Gini index as a function of $f$, for the two rules: {\it
fair} and {\it loser}.} \label{gini2}
\end{figure}

We summarize the comparison of the two rules depicting the Gini
coefficient as a function of the $f$ parameter. In Fig. \ref{gini2}
we can see clearly that the Gini index decreases when increasing the
parameter $f$, but the two rules give unequal results: the {\it
fair} rule spans on a wider interval of Gini index values than the
{\it loser} rule. The crossing of both lines at $f=0.3$ confirms
what was apparent in Fig.~\ref{hist}(a-b): for that value the same
distribution emerges with both rules. In fact most of the previous
analysis of wealth distribution can be see clearly here: while the
{\it fair} rule gives a little bit more ``humanitarian''
distributions for high $f$ values (more protective economies), the
{\it loser} rule rescue more agents from the poorest region in the
case of small values of $f$.

\section{Wealth time autocorrelation function}
We define a time autocorrelation function for the wealth as follows:
$$corr(\tau,t) = \frac{\sum_{i=1}^N w_i(\tau)w_i(\tau+t)}{\sum_{i=1}^N
w_i(\tau)w_i(\tau)}$$
\begin{figure}[htb]
\centering{\includegraphics[width=5cm,angle=-90,clip=true]{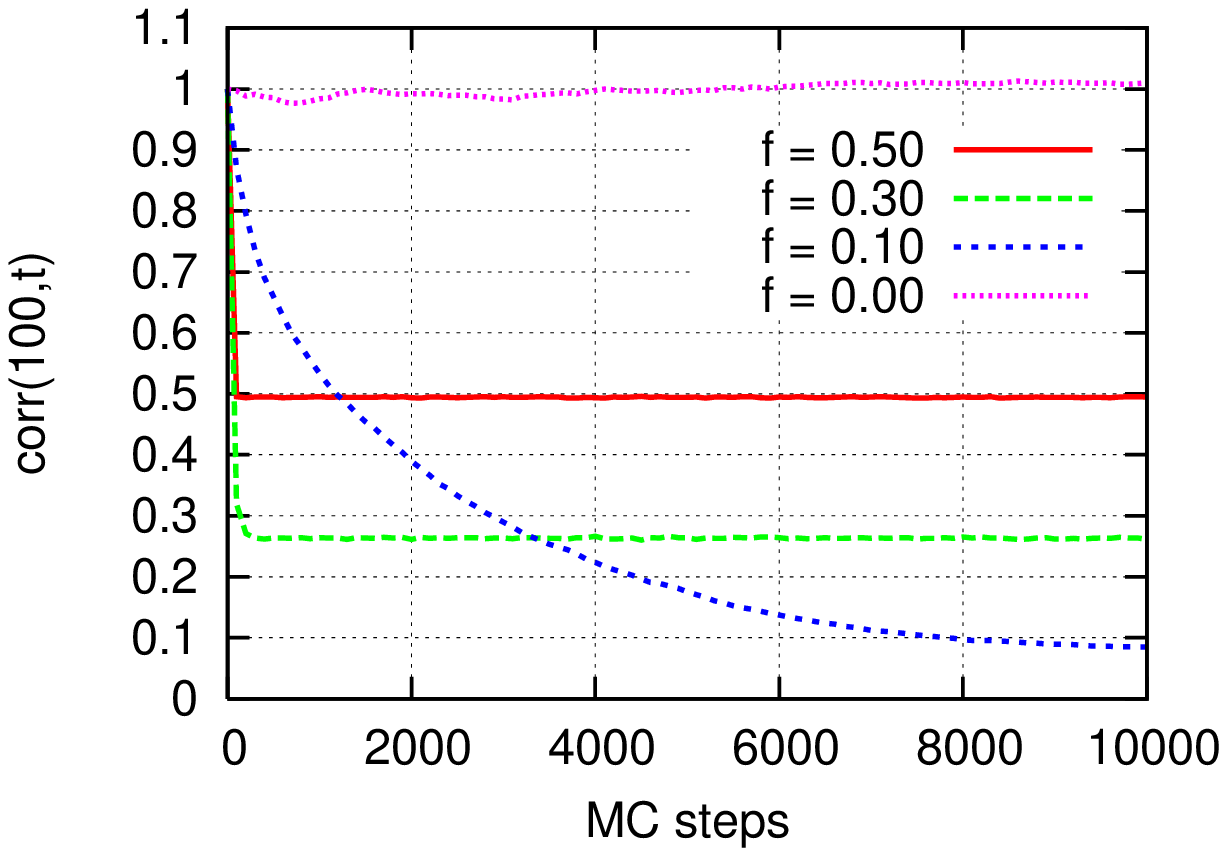}
           \includegraphics[width=5cm,angle=-90,clip=true]{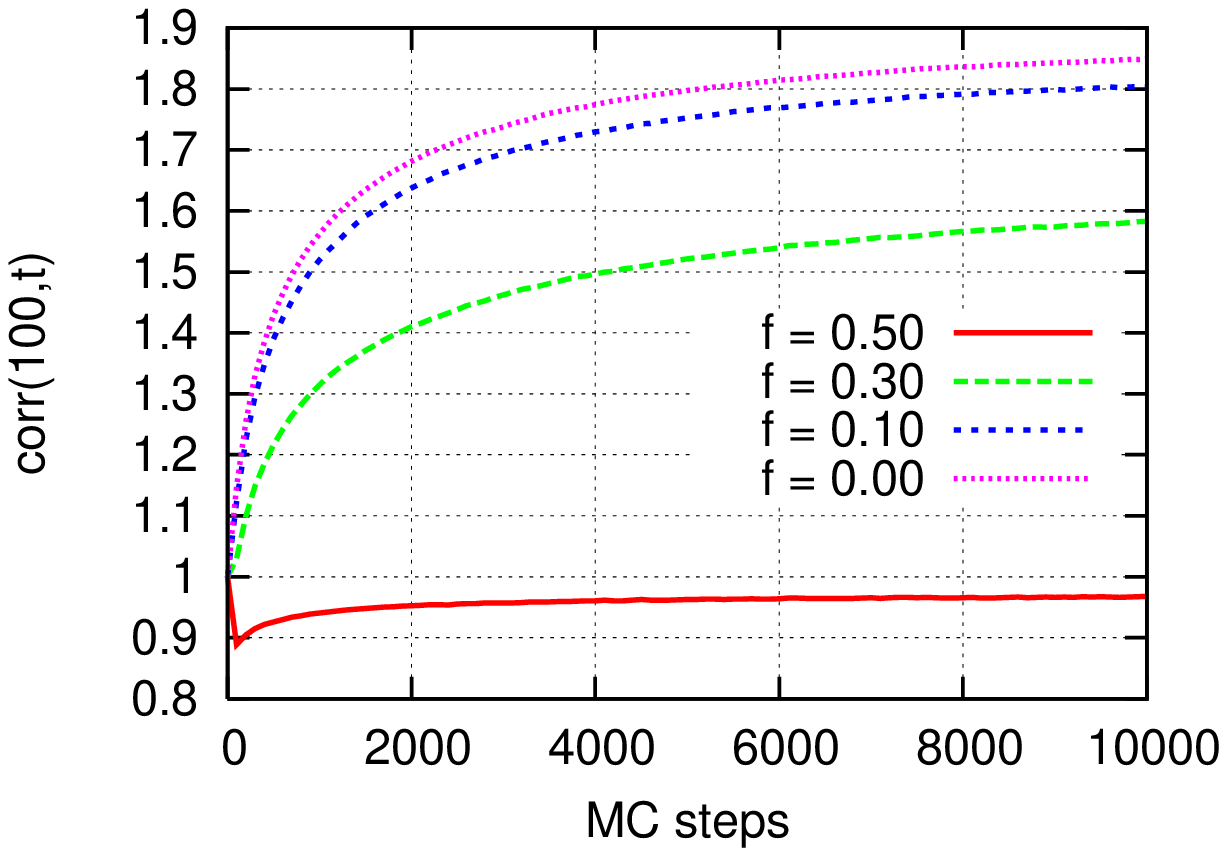}}
\caption{Wealth time autocorrelation function (defined in text) for
$\tau=100$ MC steps and different values of the parameter $f$, with
{\it fair} (left) and {\it loser} (right) rule.} \label{corr}
\end{figure}
The behavior depicted by the time correlation functions is very
rich, in a way that is not perceive neither by the distribution or
the Gini index. The behaviour in the rules are very disparate, but
in both cases clearly signs of aging are observed (we checked this
varying the initial time $\tau$). While in the {\it fair} case the
time correlation decreases, in the {\it loser} case, after an
initial and short decrease, the correlation function increases to
values bigger than one.  We remark that in the {\it fair} case and
for $f=0$ the time correlation is stable and equal to 1, because the
system condensates very fast and the changes in the wealth of the
agents are very small. But when increasing the value of $f$ the time
correlation function decreases, suggesting a higher degree of social
mobility as time evolves. On the other hand, for the {\it loser}
rule the time correlation exhibits slow dynamics (glassy behaviour,
also observed for the {\it fair} rule in the case $f=0.1$)); that
means that the system requires a large time-period to attain an
almost stable situation (as that represented in Fig.\ref{hist}), and
this characteristic time increases when increasing the size of the
system, as expected. Another interesting point is that, with the
exception of $f=0.5$, the value of the autocorrelation function is
bigger than one, indicating a kind of bias in the evolution: on
average rich agents are becoming richer and poor agents poorer. This
result, combined with the conservation of the total wealth, explains
the fact that the correlation function increases as time goes by. A
full discussion of these results, considering also different values
of $\tau$ will be presented elsewhere.

\section{Conclusions}
Our results clearly show that in many cases the system does not
arrive to a steady configuration (a remarkable exception is the case
$f=0.5$, when the probability has its strongest bias to the poorer
partner). This point is not decisive for the description of economic
systems as usually they are not in a steady state neither they are
conservative, but the implications could be interesting for possible
physical systems that behaves in a similar glassy way.

Probably the most relevant result is the fact that the {\it loser}
rule appears to produce a less unequal wealth distributions than the
one we call {\it fair} rule. That is valid for values of the $f <
0.3$, which represent situations more close to real economic
systems. Thus, the {\it loser} rule behaves  somehow like an {\it
unfair} lottery, in the sense that the richer agent risks more -- in
average -- than its poorer partner, but has less chances to win. As
a consequence, this bias attenuates the inequalities induced by low
$f$ values. It seems to us that this result is an indication that
the best way to diminish inequality does not pass only through equal
opportunity (fair rule) but through some kind of positive action
increasing the odds of poorer strata of the society.


\end{document}